\documentclass[a4paper,11pt]{article}

\usepackage{pos}
\usepackage{subfig}
\usepackage{lineno}

\newcommand{\ipred}{\ensuremath{\mathcal{I}}}

\newcommand{\zentir}{\ensuremath{\theta_{\mathrm{TIR}}}}
\newcommand{\om}{\omega}

\newcommand{\abs}[1]{\left|#1\right|}
\newcommand{\corsika}{CORSIK\kern -0.05em A 8}
\newcommand{\eisvogel}{Eisvogel}
\newcommand{\nuradiomc}{NuRadioMC}

\title{Observation of Broadband In-ice Radiation from Impacting High-Energy Cosmic Rays}
\ShortTitle{In-ice Radiation from Cosmic Rays}

\author*[a]{Philipp Windischhofer} 
\author[a]{Nathaniel Alden}
\onbehalf{for the ARA Collaboration \\{\normalsize \normalfont(a complete list of authors can be found at the end of the proceedings)}\\}

\affiliation[a]{Dept. of Physics, Enrico Fermi Institute,
  Kavli Institute for Cosmological Physics,\\
  University of Chicago, Chicago, IL 60637}

\emailAdd{windischhofer@uchicago.edu}

\abstract{
  We present the first experimental evidence for in-ice radiofrequency emission from high-energy particle cascades developing
  in the Antarctic ice sheet.
  In 208 days of data recorded with the phased-array trigger of the Askaryan Radio Array, we detect 13 events with
  impulsive radiofrequency pulses originating from below the ice surface.
  Considering only the arrival angles and timing properties, this rate is inconsistent with an a-posteriori background expectation 
  for thermal noise events and on-surface events at the level of $3.5\,\sigma$, which
  rises to $5.1\,\sigma$ when additionally considering impulsivity.
  The observed event geometry, event rate, signal shape, spectral content, and electric field polarization
  are consistent with Askaryan radiation from cosmic ray air shower cores impacting the ice sheet.
  For the brightest events, the angular radiation pattern independently favors an extended cascade-like emitter over
  a pointlike source.
}

\ConferenceLogo{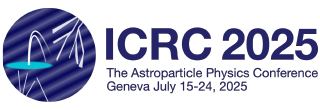}

\FullConference{39th International Cosmic Ray Conference (ICRC2025)\\
 15–24 July 2025\\
Geneva, Switzerland\\}

\begin{document}

\maketitle

\section{Introduction}
\label{sec:introduction}
\noindent
A high-energy electromagnetic cascade developing in normal matter generates an excess of negative charge:
positrons continuously annihilate, and Compton, Bhabha, and M{\o}ller scattering draws electrons from the surrounding material
into the shower.
The net-negative charge distribution formed by the particles in the shower front moves at super-luminal
speeds in the medium and, for cascades developing in a dielectric, emits coherent Cherenkov-like
electromagnetic radiation.
Initially predicted by Askaryan in 1962 \cite{askaryan_1}, measurements at particle accelerators in the 
2000s \cite{sand_askaryan, salt_askaryan, polyethylene_askaryan, ice_askaryan} have observed and 
studied this radiofrequency (RF) signature from particle showers in dense dielectrics.
Outside the laboratory, in-air Askaryan emission from cosmic ray-induced extensive air showers has been experimentally
identified in the 2010s \cite{Auger_air_askaryan, LOFAR_air_askaryan, CODALEMA_air_askaryan}.
Here, the Askaryan effect contributes only about 10\% of the overall electric field
amplitude \cite{LOFAR_air_askaryan, CODALEMA_air_askaryan} and emission from the geomagnetic separation of charges 
is dominant \cite{geomag_theory_1, geomag_theory_2}.

\begin{figure}[tb]
  \centering
  \subfloat[][]{\includegraphics[width=0.47\textwidth]{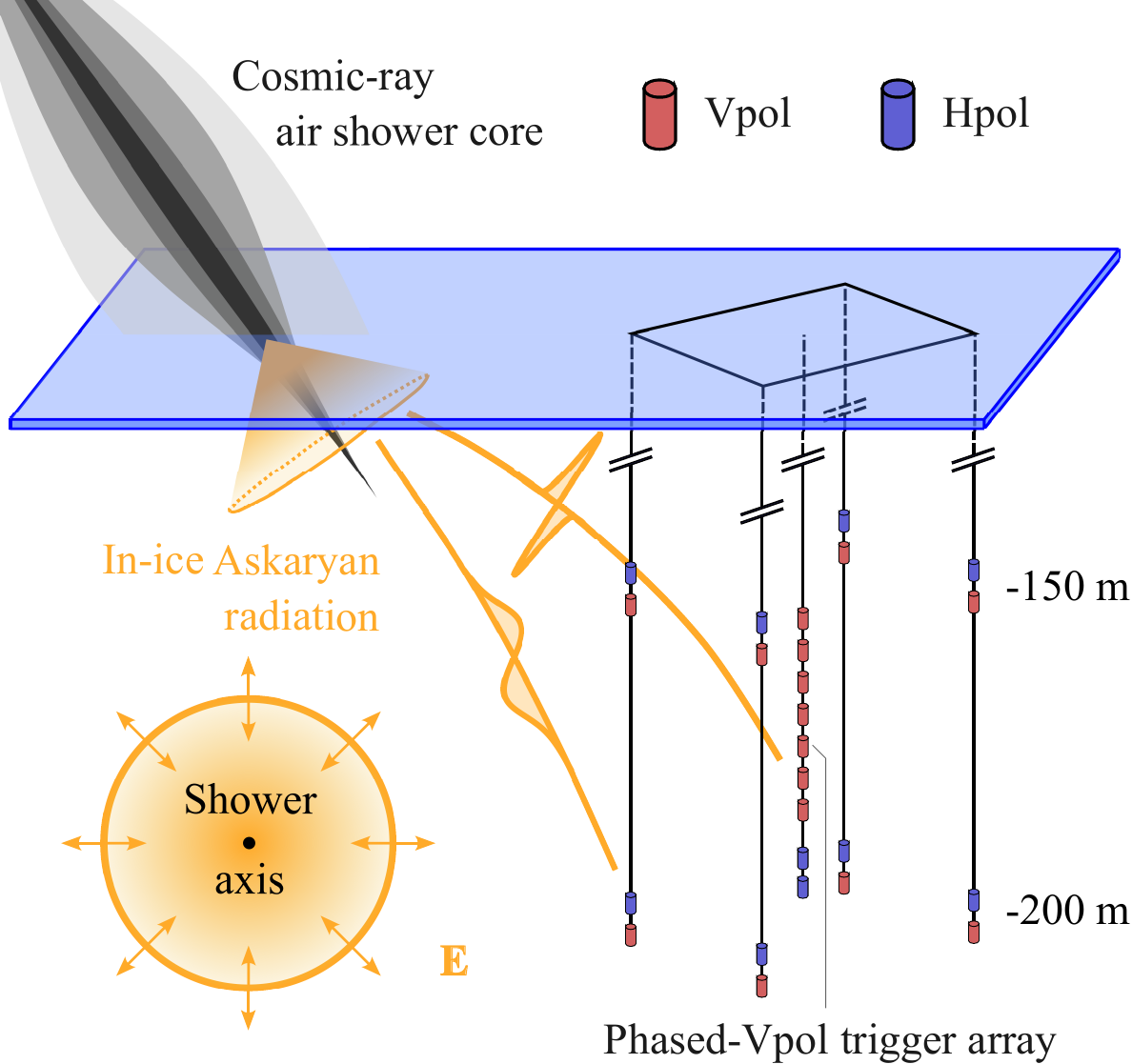}\label{fig:detector}}\quad
  \subfloat[][]{\includegraphics[width=0.49\textwidth]{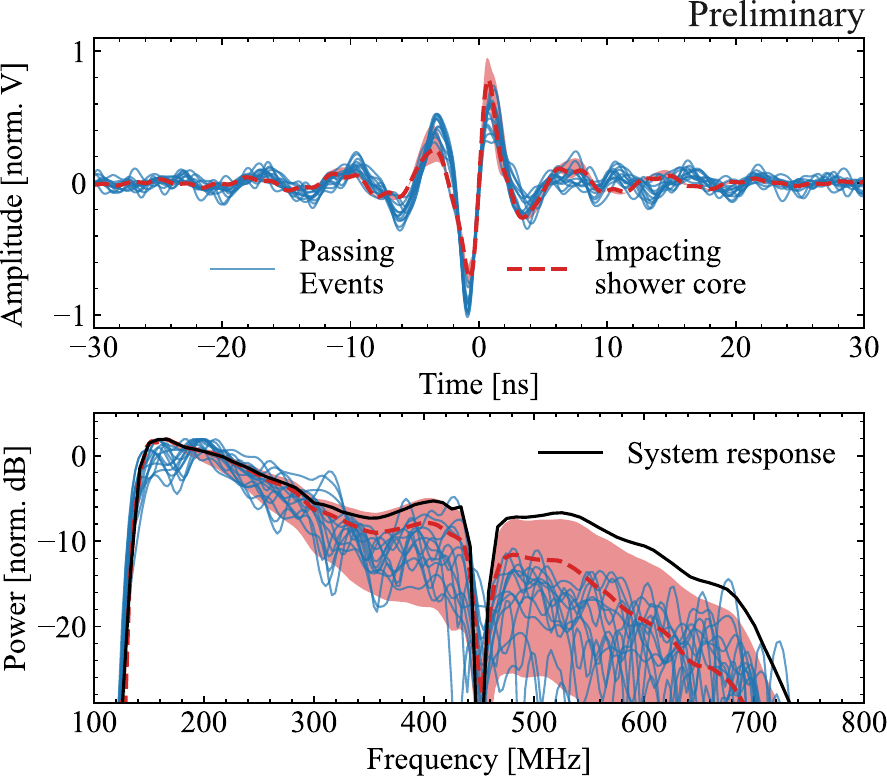}\label{fig:signals}}
  \caption{
    \protect\subref{fig:detector}
    The dense cores of cosmic ray air showers (black) emit coherent Askaryan radiation in the near-surface ice.
    This radiation has a characteristic radial electric field polarization with respect to the shower axis (bottom left)
    and is strongly beamed around the Cherenkov angle, leading to differences in spectral content when observed
    in multiple A5 receivers.
    \protect\subref{fig:signals}
    Top: VPol waveforms of the 13 passing events measured by the phased array, normalized to their peak-to-peak
    amplitudes.
    Bottom: Relative power density spectra compared with the instrument response (black).
    In both panels, simulated signals are shown in red.
  }
\end{figure}

The Askaryan Radio Array (ARA) \cite{ARA_instrument_paper}, near South Pole, seeks to employ the Askaryan effect to search for
ultrahigh-energy ($\gtrsim100\,\mathrm{PeV}$) astrophysical neutrinos interacting in the
Antarctic ice sheet.
ARA consists of five independent stations, each spaced two kilometers apart.
Each station consists of a total of eight birdcage-style vertically-polarized dipole antennas (VPol) and eight
ferrite-loaded quad-slot antennas (HPol) deployed as strings in four air-filled boreholes at up to 200\,m depth.
ARA Station 5 (A5), used for the study presented here and shown in Fig.~\ref{fig:detector}, is additionally equipped with
seven tightly-spaced VPol antennas and two additional HPol antennas positioned in a fifth string, forming a low-threshold
phased-VPol trigger array \cite{PA_instrument_paper}.
The antennas on the four outer strings are here used to reconstruct event arrival directions.

Cosmic rays (CRs) are a physics background to the neutrino signature in ARA: the dense cores of near-vertical CR air
showers can reach the surface of the Antarctic Plateau (at $\sim2800$\,m elevation at South Pole) and deposit a
sizable fraction ($\sim{50}\%$) of the primary energy in the ice~\cite{coleman2024iniceaskaryanemissionair},
generating RF transition radiation (TR) at the air-ice interface \cite{transition_radiation1} and Askaryan radiation in the
top 5--10\,m of the ice sheet \cite{Original_faerie, FAERIE, De_Kockere_2024, shower_sim_thesis}.
Recent calculations \cite{coleman2024iniceaskaryanemissionair} predict Askaryan event rates of $10$--$100\,\mathrm{yr}^{-1}$
for an ARA-like detector station, larger than the expected neutrino detection rate by at least two orders of
magnitude \cite{neutrino_rate}.
This underlines the importance of a detailed characterization of this process, but also suggests that it can be used
to study the in-ice Askaryan signature \textsl{in-situ} to validate the detection mechanism and simulation approaches
used in a neutrino search.

\section{Event Selection}
\noindent
Motivated by these considerations, we here examine events that have previously been identified in a sideband of a neutrino
search \cite{ara-lowthresh-analysis}.
The search was conducted using 208 days of data recorded with the phased-VPol trigger array installed at A5 in the period
from February--November 2019.
The selection criterion is based on a multivariate discriminant trained to identify Askaryan emission from simulated
neutrino events, and the neutrino search region was defined, among other cuts, by requiring the signal arrival zenith
angle $\theta$ measured at the phased array to be greater than $57^\circ$ in order to avoid contamination from sources
near or above the surface.
For reference, sources located on or above the surface and producing planar wave fronts populate the region
$\theta < \zentir{}$, with the angle of total internal reflection (TIR) given by $\zentir{} \sim 34.5^\circ$.

Following unblinding, 46 events were observed in the region $\theta \leq 57^\circ$, orthogonal to the neutrino search region.
Here, we focus on events in the range $38^\circ \leq \theta \leq 57^\circ$, in which radiation arrives at the phased array
from directions shallower than the TIR angle and (background) contributions from distant in-air and on-surface sources
are suppressed.
A total of 13 events are observed in this region.
These events are impulsive, isolated in time, occur at a uniform rate of $22.9^{+8.2}_{-6.2}\,\mathrm{yr}^{-1}$, show no
periodicity with calendar day, UTC hour, minute, or second, and are consistent with a uniform distribution in azimuth
(Kuiper's test yields a $p$-value of 0.92).
Twelve events occur during polar night, consistent with the livetime fraction, when human activity at the nearby
South Pole Station ($\sim 5\,\mathrm{km}$ away from A5) is at a minimum and activity off-station is limited.
No correlation has been identified with periods of high winds, which are known to be able to produce impulsive radio
emission via the triboelectric effect \cite{triboelectric}.
The signal-to-noise ratio\footnote{The signal-to-noise ratio is defined as the ratio of the peak-to-peak
signal amplitude and the noise RMS amplitude.} (SNR), averaged over the phased-VPol receivers,
ranges from 4.4 to 18.2, with three events showing an SNR above 15.

\section{Background Estimate}
\label{sec:background_estimate}
\noindent
To compare the observed event rate with expected non-physics processes, we consider three distinct classes of backgrounds,
briefly summarized below.
The contribution from thermal noise triggers in the signal region is estimated at $0.14 \pm 0.05$ expected events using the
procedure from Ref.~\cite{ara-lowthresh-analysis}.
Anthropogenic signals from distant near-horizon sources leaking into the signal region are extrapolated from
time-clustering events in a zenith control region centered around the TIR angle and estimated at $0.16^{+0.08}_{-0.06}$ expected events.
Finally, signals from on-surface sources within a radius of $\sim250$\,m around A5 can efficiently couple into the ice and enter the
signal region directly. 
Using control samples enriched in known on-surface activity (e.g.~a GNSS-equipped snowmobile passing by A5) we find that such processes typically 
produce time-clustering events of which only a small fraction ($\lesssim$10\%) reaches impulsivity values% 
\footnote{We quantify the impulsivity of an event by the ratio of maximum instantaneous signal power to mean signal power
calculated in the 50\% of the trace closest to the peak.
Impulsivity defined as such is nonnegative and unbounded from above; restricting the calculation of the mean to a fraction
of the full trace length reduces its correlation with signal SNR.}
comparable to the passing events.
Assuming these observations to be representative, we find
a Feldman-Cousins upper limit of 0.12 expected events at 95\% confidence level for the population
of time-isolated on-surface background events with impulsivity similar to the passing events.
The observed event yield represents a deviation from the combined background expectation at an \textsl{a-posteriori} 
significance of $5.1\,\sigma$,
determined from the distribution of the profile-likelihood ratio test statistic \cite{pdg-statistics} under the background-only hypothesis.
In a separate analysis where no impulsivity information is considered, the significance is $3.5\,\sigma$.

\section{Askaryan Event Rate}
\label{sec:event_rate}
\noindent
To estimate the Askaryan event rate expected from impacting air shower cores, we follow the approach of
Ref.~\cite{coleman2024iniceaskaryanemissionair}.
Briefly, we use \nuradiomc{} \cite{NuRadioMC} and a parameterized Askaryan emission model \cite{ARZ} derived for neutrino-induced
cascades in ice to efficiently simulate a large number of impacting shower-core-like events.
We weight these events by the CR energy flux at the ice surface and correct for differences in the overall emission strength
with a scale factor calibrated against microscopic simulations of impacting shower cores (described below).
Absolute yield estimates carry sizable uncertainties as a result, but relative yields and distributions are robustly simulated.
We predict a rate of $10$--$45\,\mathrm{yr}^{-1}$ in the signal region of interest, consistent with the observed rate.
The simulated rate peaks for CRs depositing $\sim10^{16.8}$\,eV in the ice.
We use the same approach to simulate the distributions of signal arrival direction and polarization angle, shown below.

\section{Event Properties}
\label{sec:signal_properties}
\noindent
We next turn to a closer examination of the passing events, considering additional observables.

\paragraph{Signal shape}

Fig.~\ref{fig:signals} shows the VPol waveforms of all passing events, with the instrumental phase response removed and the
time-aligned waveforms from all phased-VPol receivers combined to improve the SNR.
All waveforms contain a single pulse with a common polarity
and qualitatively similar shape, suggesting a common emission mechanism.
The bottom panel of Fig.~\ref{fig:signals} displays the power spectra of the (upsampled, time-windowed) waveforms in comparison
with the instrumental response, illustrating their broadband and band-filling nature.
We find agreement of both the phase and spectral shape with microscopic simulations of the in-ice
radio signature generated by an ice-impacting CR air shower, in Fig.~\ref{fig:signals} shown for a vertical shower induced by a
proton primary with an energy of $10^{17}\,\mathrm{eV}$.
The red envelope indicates a range of $\sim3^\circ$ near the approximate in-ice Cherenkov angle, representative
for events entering the signal region.
We use \corsika{} \cite{corsika8} to simulate the cross-media particle cascade and \eisvogel{} \cite{eisvogel, eisvogel_arena} for
fully-electrodynamic radio emission and propagation, which includes TR at the air-ice surface and Askaryan
radiation from the in-ice cascade.
These simulations indicate that the excess charge of the dense shower core at impact is below 10\%--20\% of the maximum in-ice charge
excess.
A similar charge-excess ratio applies in measurements conducted in the GeV electron beam at SLAC \cite{ice_askaryan}, where TR at
an air-ice interface was found to be subdominant.
We expect the same conclusion to hold here, where the reduced sharpness of the dielectric boundary between air and snow
is expected to further reduce the contribution from TR relative to in-ice Askaryan radiation \cite{transition_radiation1}.

\paragraph{Signal arrival directions}

\begin{figure}
  \centering
  \raisebox{3mm}{\subfloat[][]{\includegraphics[width=0.48\textwidth]{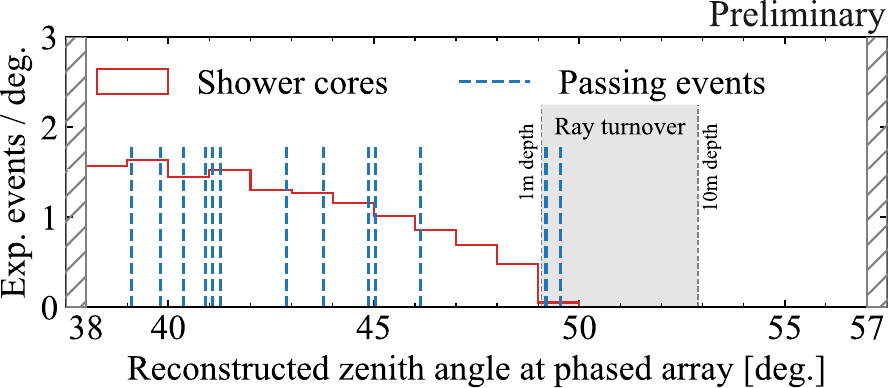}\label{fig:zenith}}}\quad
  \subfloat[][]{\includegraphics[width=0.48\textwidth]{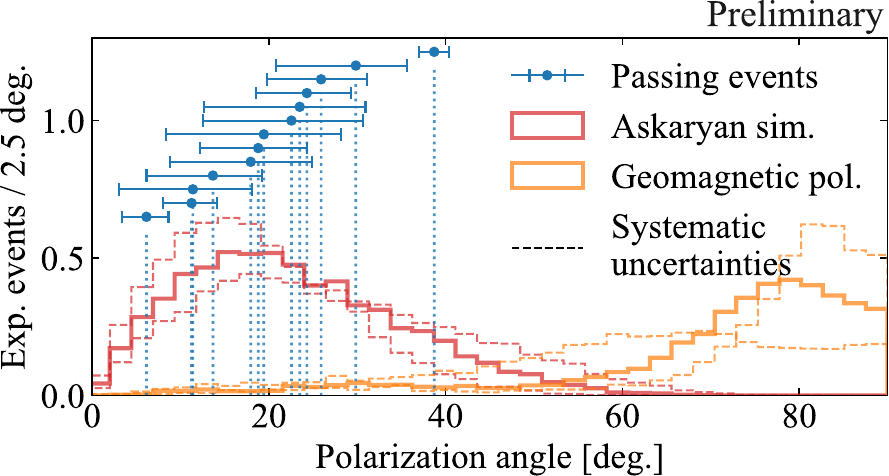}\label{fig:polarization}}
  \caption{
    \protect\subref{fig:zenith}
    Distribution of arrival zenith directions measured at the phased array for the events in the signal region
    ($38^\circ$--$57^\circ$)
    and compared with the simulated distribution for in-ice Askaryan emission from impacting shower cores (normalized to the observed yield).
    The gray region indicates the minimum depth of a source such that its radiation reaches the phased array at the given
    angle.
    \protect\subref{fig:polarization}
    Distribution of measured polarization angles for the passing events; the error bars indicate statistical uncertainties.
    The solid red distribution shows the simulation prediction for in-ice Askaryan radiation from impacting shower cores
    (normalized to the observed yield).
    The orange distributions show geomagnetic polarization at the phased array, given the observed VPol signals.
    The dashed distributions bound experimental systematic uncertainties.    
  }
\end{figure}

The distribution of signal arrival direction measured at the phased array carries information about the radial location of
the source and its emission pattern.
Fig.~\ref{fig:zenith} compares the arrival zenith angles reconstructed for the passing events with the expectation for
Askaryan radiation from impacting shower cores.
The Kolmogorov-Smirnov (KS) test yields a $p$-value of 0.8, indicating good agreement.
Notably, no events are observed for zenith angles $\gtrsim 50^\circ$, where ray bending in the inhomogeneous permittivity
distribution of the ice prevents radiation originating near the surface from reaching the detector, independently suggesting
a near-surface source.
Due to refraction, TR peaking close to the shower axis instead of at the Cherenkov angle can reconstruct
into our signal region only for highly-inclined showers with zenith angles greater than $\sim 55^\circ$.
The impacting flux in this range is low, accounting for at most 0.5\% of the total flux above $10^{17}\,\mathrm{eV}$
and suppressing the possible TR event rate relative to the Askaryan rate.

\paragraph{Polarization}

Fig.~\ref{fig:polarization} compares the electric field polarization angles measured for the passing events with the expectation
for in-ice Askaryan radiation from air shower cores.
The polarization analysis uses the HPol antenna pair at the bottom of the phased-array string and the two lowermost phased-VPol
antennas, all co-located within 6\,m.
To define the polarization angle, we follow Ref.~\cite{aera-polarization-reconstruction} and first measure the noise-subtracted
energy fluences $f_\phi$ and $f_\theta$ for azimuthal and polar field polarizations, using time-aligned and summed waveforms from
the HPol and VPol antenna pairs, respectively.
The fluences are constructed as $f_{\phi, \theta} = \int dt\, E^2_{\phi, \theta}(t) - c_{\phi, \theta}$, where the integration defines
a time-window centered on the pulse and $c_{\phi, \theta}$ is constructed such that the fluence vanishes in the mean when
evaluated on events containing no signal.
The measurement is performed in the 170\,MHz--400\,MHz band, where both antenna types show good response \cite{ARA_instrument_paper}.
We then use the (absolute) angle $\abs{\psi} = \arctan \sqrt{\abs{f_\phi} / \abs{f_\theta}}$ to characterize the electric field polarization,
where small angles indicate a mostly-$E_\theta$ polarization.
Experimental systematic uncertainties on the simulated distributions are dominated by uncertainties on the in-borehole HPol antenna response.
The effect of a $\pm 50\%$ variation of the HPol antenna effective length is shown by the dashed distributions in Fig.~\ref{fig:polarization}.
The data are however well-compatible with the nominal prediction at a KS $p$-value of 0.5.
Although we do not directly reconstruct the shower axis, simulation predicts the signal region to select showers with typical inclinations
in the range $30^\circ$--$40^\circ$.
Askaryan radiation is radially polarized with respect to the shower axis, and inclined showers in this range can generate events with
significant $E_\phi$ content, cf.~Fig.~\ref{fig:detector}.

The measured polarization is, however, inconsistent with in-air geomagnetic radiation (also shown in Fig.~\ref{fig:polarization}), which
has a characteristic $\mathbf{v} \times \mathbf{B}$ polarization in the plane perpendicular to the shower axis.
At South Pole, the local geomagnetic dip angle is $\sim{72}^\circ$ \cite{wmm}, leading to an $E_\phi$-dominated radiation signature and a
polarization angle distribution significantly different from the observed
(with KS $p$-value $2.3\times10^{-14}$, a $7.5\,\sigma$ deviation).

\paragraph{Test for a Shower-like Source}

The view directions of the A5 receivers vary by $10^\circ$--$15^\circ$ for a near-surface source, across which the beamed
in-ice Askaryan radiation pattern is expected to produce detectable differences in the absolute radiation intensity and spectral content.
Below, we describe a procedure using signals from all receivers to identify this signature in data, directly probing
the spatial structure of the radiation source in a manner independent of the results presented thus far.
To illustrate the method, we first consider a simple model of an in-ice cascade, in which a variable point charge $q(z)$ moving at the
speed of light along the $z$-axis represents the charge excess.
The emitted electric field amplitude at distance $R$ and angular frequency $\om$ scales as \cite{diff_prd}
\begin{equation}
  |R\, \mathbf{E}(\om, \theta)| \sim \mu_0 \om \sin\theta \int dz\, q(z) e^{i z \frac{n \om}{c} \left(\cos\theta_c-\cos\theta\right)},
  \label{eq:emission_shower}
\end{equation}
where $\theta$ is the angle with respect to the shower axis at which the source is observed, $\theta_c$ is the Cherenkov angle,
$n$ is the refractive index of the medium, and $\mu_0$ is the vacuum permeability.
The Fourier integral in Eq.~\ref{eq:emission_shower} produces the signature characteristic of diffraction phenomena,
where the angular width of the radiation intensity beam pattern $\ipred{}(\om, \theta)\sim|\mathbf{E}(\omega, \theta)|^2$ is 
inversely proportional to the frequency.
This implies that, as $\theta$ is varied, the rate of change in the received signal intensity $\ipred{}$ scales proportionally
with frequency.
For $\om_2 > \om_1$,
\begin{equation}
  \frac{d\log \ipred{}(\om_2, \theta)}{d\theta} \biggr/ \frac{d\log \ipred{}(\om_1, \theta)}{d\theta} =
  \frac{d\log \ipred{}(\om_2)}{d\log \ipred{}(\om_1)} > 1.
  \label{eq:hflf_slope}
\end{equation}
This is in contrast to pointlike (non-diffractive) sources, such as electrically-small transmitters or TR
at the air-ice interface, for which the angular radiation pattern is (approximately) independent of
frequency \cite{antenna_theory, transition_radiation1, transition_radiation2, transition_radiation3}, so that
$d\log \ipred{}(\om_2) / d\log \ipred{}(\om_1) = 1$.

\begin{figure}
  \includegraphics[width=\textwidth]{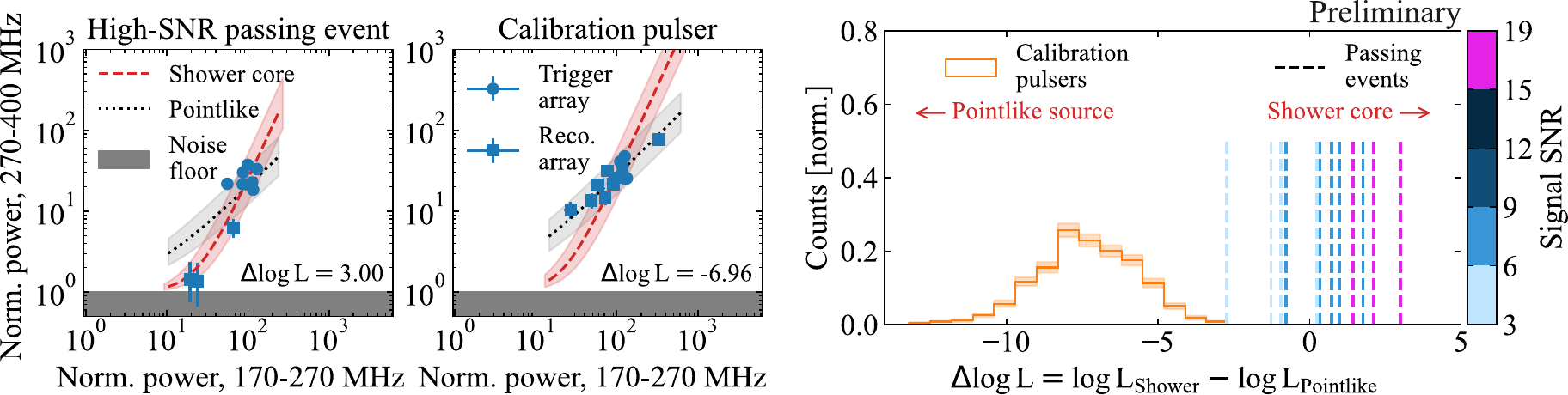}
  \caption{
    Left: Noise-normalized VPol signal intensities in two frequency bands, for an example signal
    candidate event and a calibration transmitter event.
    Model predictions are overlaid for impacting shower
    cores (dashed) and a pointlike source (dotted).
    Data uncertainties only include statistical components, while experimental systematic uncertainties are indicated by the colored envelopes.
    The respective $\Delta \log L$ values (defined in the main text) are also shown.
    Right: Distributions of $\Delta \log L$ for the candidate signal events (vertical lines, with SNR indicated by color) and a sample of calibration
    pulser events.
    }
  \label{fig:askaryan}
\end{figure}

We perform a per-event likelihood-ratio test to discriminate between these two scenarios.
The test is conducted in terms of the (time-windowed) VPol signal intensities as measured at the digitizer in a higher-frequency (HF) band,
270\,MHz--400\,MHz, and a lower-frequency (LF) band, 170\,MHz--270\,MHz, normalized by the average recorded noise intensity (from minimum-bias triggers)
to correct for amplifier gain differences across channels.
For impacting shower cores, simulation shows that the logarithmic signal intensities in these two bands are related in good approximation as
$\log \ipred{}_\mathrm{HF} \sim 2.2 \log \ipred{}_\mathrm{LF}$ within $\sim7^\circ$ of the Cherenkov angle, while absolute radiation intensities
vary by a factor 10--100.
The test statistic is then defined as $\Delta \log L = \log L_{\mathrm{Shower}} - \log L_{\mathrm{Pointlike}}$, where the shower
likelihood uses the above-mentioned parameterization, and the likelihood under the null hypothesis $L_{\mathrm{Pointlike}}$
assumes $\log \ipred{}_\mathrm{HF} \sim \log \ipred{}_\mathrm{LF}$. 
It is constructed symmetrically such that $\Delta \log L \approx 0$ for events not showing clear evidence for either hypothesis.
Note that differences in the antenna beam pattern or the in-ice signal propagation distance between receivers are not corrected for and do not
enter the likelihood; their impact on the intensity is estimated as 20\% each, much smaller than the targeted effect.
As defined, the test statistic is agnostic to the event geometry, enabling the analysis of events with insufficient
information for a shower-axis reconstruction.
The power of the test is, however, greatest for high-SNR events which illuminate many channels.

Fig.~\ref{fig:askaryan} (left) shows the signal intensities in the two bands for a high-SNR passing event.
There are visible differences in the spectral shape across receivers, consistent with the expectation of beamed radiation from
an impacting shower core.
This is in contrast to signals from a nanosecond-pulsed calibration transmitter, also shown, which are assigned negative test statistic values and
identified as originating from a pointlike source.
The $\Delta\log L$ values of all passing events are compared in Fig.~\ref{fig:askaryan} (right).
Seven of the eight highest-SNR events have positive $\Delta\log L$ and the three brightest events show $\Delta\log L > 1$, consistently
indicating a preference for a shower-like source.

\section{Conclusions}
\label{sec:conclusions}
\noindent
We find that the observed rate of impulsive near-surface events forms an excess over the background expectation with a significance 
of $5.1\,\sigma$ when signal impulsivity information is considered.
The rate, signal waveforms, spectral content, arrival direction and electric field polarization of these events are consistent with expectations
for in-ice Askaryan radiation from ice-impacting air shower cores.
For the brightest observed events, the variation of the spectral content across nearby receivers separately indicates a strongly-beamed radiation
pattern consistent with an extended in-ice shower core.
In-air geomagnetic emission is disfavored by the event geometry and the signal polarization, while TR is suppressed by the relatively small
shower-core charge excess at the surface
and disfavored by the observed arrival direction distribution in the signal region.
We expect that analyses of larger datasets and extending to the above-TIR zenith range may yield TR event candidates in the future.

Given the observed rate, the full A5 dataset should already contain more than hundred events similar to those shown here.
Their neutrino-like signature presents a valuable target to validate detailed simulation models
\cite{cosmin-fdtd, arianna-surface-mode} and optimize methods to reject these non-neutrino physics backgrounds.

\bibliographystyle{ICRC}
\renewcommand*{\bibfont}{\footnotesize}
\setlength{\bibsep}{-0.03em}
\bibliography{references}

\clearpage

% ICRC list for ARA Collaboration
\section*{Full Author List: ARA Collaboration (June 30, 2025)}

\noindent
N.~Alden\textsuperscript{1}, 
S.~Ali\textsuperscript{2}, 
P.~Allison\textsuperscript{3}, 
S.~Archambault\textsuperscript{4}, 
J.J.~Beatty\textsuperscript{3}, 
D.Z.~Besson\textsuperscript{2}, 
A.~Bishop\textsuperscript{5}, 
P.~Chen\textsuperscript{6}, 
Y.C.~Chen\textsuperscript{6}, 
Y.-C.~Chen\textsuperscript{6}, 
S.~Chiche\textsuperscript{7}, 
B.A.~Clark\textsuperscript{8}, 
A.~Connolly\textsuperscript{3}, 
K.~Couberly\textsuperscript{2}, 
L.~Cremonesi\textsuperscript{9}, 
A.~Cummings\textsuperscript{10,11,12}, 
P.~Dasgupta\textsuperscript{3}, 
R.~Debolt\textsuperscript{3}, 
S.~de~Kockere\textsuperscript{13}, 
K.D.~de~Vries\textsuperscript{13}, 
C.~Deaconu\textsuperscript{1}, 
M.A.~DuVernois\textsuperscript{5}, 
J.~Flaherty\textsuperscript{3}, 
E.~Friedman\textsuperscript{8}, 
R.~Gaior\textsuperscript{4}, 
P.~Giri\textsuperscript{14}, 
J.~Hanson\textsuperscript{15}, 
N.~Harty\textsuperscript{16}, 
K.D.~Hoffman\textsuperscript{8}, 
M.-H.~Huang\textsuperscript{6,17}, 
K.~Hughes\textsuperscript{3}, 
A.~Ishihara\textsuperscript{4}, 
A.~Karle\textsuperscript{5}, 
J.L.~Kelley\textsuperscript{5}, 
K.-C.~Kim\textsuperscript{8}, 
M.-C.~Kim\textsuperscript{4}, 
I.~Kravchenko\textsuperscript{14}, 
R.~Krebs\textsuperscript{10,11}, 
C.Y.~Kuo\textsuperscript{6}, 
K.~Kurusu\textsuperscript{4}, 
U.A.~Latif\textsuperscript{13}, 
C.H.~Liu\textsuperscript{14}, 
T.C.~Liu\textsuperscript{6,18}, 
W.~Luszczak\textsuperscript{3}, 
A.~Machtay\textsuperscript{3}, 
K.~Mase\textsuperscript{4}, 
M.S.~Muzio\textsuperscript{5,10,11,12}, 
J.~Nam\textsuperscript{6}, 
R.J.~Nichol\textsuperscript{9}, 
A.~Novikov\textsuperscript{16}, 
A.~Nozdrina\textsuperscript{3}, 
E.~Oberla\textsuperscript{1}, 
C.W.~Pai\textsuperscript{6}, 
Y.~Pan\textsuperscript{16}, 
C.~Pfendner\textsuperscript{19}, 
N.~Punsuebsay\textsuperscript{16}, 
J.~Roth\textsuperscript{16}, 
A.~Salcedo-Gomez\textsuperscript{3}, 
D.~Seckel\textsuperscript{16}, 
M.F.H.~Seikh\textsuperscript{2}, 
Y.-S.~Shiao\textsuperscript{6,20}, 
S.C.~Su\textsuperscript{6}, 
S.~Toscano\textsuperscript{7}, 
J.~Torres\textsuperscript{3}, 
J.~Touart\textsuperscript{8}, 
N.~van~Eijndhoven\textsuperscript{13}, 
A.~Vieregg\textsuperscript{1}, 
M.~Vilarino~Fostier\textsuperscript{7}, 
M.-Z.~Wang\textsuperscript{6}, 
S.-H.~Wang\textsuperscript{6}, 
P.~Windischhofer\textsuperscript{1}, 
S.A.~Wissel\textsuperscript{10,11,12}, 
C.~Xie\textsuperscript{9}, 
S.~Yoshida\textsuperscript{4}, 
R.~Young\textsuperscript{2}
\\
\\
\textsuperscript{1} Dept. of Physics, Enrico Fermi Institute, Kavli Institute for Cosmological Physics, University of Chicago, Chicago, IL 60637\\
\textsuperscript{2} Dept. of Physics and Astronomy, University of Kansas, Lawrence, KS 66045\\
\textsuperscript{3} Dept. of Physics, Center for Cosmology and AstroParticle Physics, The Ohio State University, Columbus, OH 43210\\
\textsuperscript{4} Dept. of Physics, Chiba University, Chiba, Japan\\
\textsuperscript{5} Dept. of Physics, University of Wisconsin-Madison, Madison,  WI 53706\\
\textsuperscript{6} Dept. of Physics, Grad. Inst. of Astrophys., Leung Center for Cosmology and Particle Astrophysics, National Taiwan University, Taipei, Taiwan\\
\textsuperscript{7} Universite Libre de Bruxelles, Science Faculty CP230, B-1050 Brussels, Belgium\\
\textsuperscript{8} Dept. of Physics, University of Maryland, College Park, MD 20742\\
\textsuperscript{9} Dept. of Physics and Astronomy, University College London, London, United Kingdom\\
\textsuperscript{10} Center for Multi-Messenger Astrophysics, Institute for Gravitation and the Cosmos, Pennsylvania State University, University Park, PA 16802\\
\textsuperscript{11} Dept. of Physics, Pennsylvania State University, University Park, PA 16802\\
\textsuperscript{12} Dept. of Astronomy and Astrophysics, Pennsylvania State University, University Park, PA 16802\\
\textsuperscript{13} Vrije Universiteit Brussel, Brussels, Belgium\\
\textsuperscript{14} Dept. of Physics and Astronomy, University of Nebraska, Lincoln, Nebraska 68588\\
\textsuperscript{15} Dept. Physics and Astronomy, Whittier College, Whittier, CA 90602\\
\textsuperscript{16} Dept. of Physics, University of Delaware, Newark, DE 19716\\
\textsuperscript{17} Dept. of Energy Engineering, National United University, Miaoli, Taiwan\\
\textsuperscript{18} Dept. of Applied Physics, National Pingtung University, Pingtung City, Pingtung County 900393, Taiwan\\
\textsuperscript{19} Dept. of Physics and Astronomy, Denison University, Granville, Ohio 43023\\
\textsuperscript{20} National Nano Device Laboratories, Hsinchu 300, Taiwan\\

\section*{Acknowledgements}

\noindent
The ARA Collaboration is grateful to support from the National Science Foundation through Award 2013134.
The ARA Collaboration
designed, constructed, and now operates the ARA detectors. We would like to thank IceCube, and specifically the winterovers for the support in operating the
detector. Data processing and calibration, Monte Carlo
simulations of the detector and of theoretical models
and data analyses were performed by a large number
of collaboration members, who also discussed and approved the scientific results presented here. We are
thankful to Antarctic Support Contractor staff, a Leidos unit 
for field support and enabling our work on the harshest continent. We thank the National Science Foundation (NSF) Office of Polar Programs and
Physics Division for funding support. We further thank
the Taiwan National Science Councils Vanguard Program NSC 92-2628-M-002-09 and the Belgian F.R.S.-
FNRS Grant 4.4508.01 and FWO. 
K. Hughes thanks the NSF for
support through the Graduate Research Fellowship Program Award DGE-1746045. A. Connolly thanks the NSF for
Award 1806923 and 2209588, and also acknowledges the Ohio Supercomputer Center. S. A. Wissel thanks the NSF for support through CAREER Award 2033500.
A. Vieregg thanks the Sloan Foundation and the Research Corporation for Science Advancement, the Research Computing Center and the Kavli Institute for Cosmological Physics at the University of Chicago for the resources they provided. R. Nichol thanks the Leverhulme
Trust for their support. K.D. de Vries is supported by
European Research Council under the European Unions
Horizon research and innovation program (grant agreement 763 No 805486). D. Besson, I. Kravchenko, and D. Seckel thank the NSF for support through the IceCube EPSCoR Initiative (Award ID 2019597). M.S. Muzio thanks the NSF for support through the MPS-Ascend Postdoctoral Fellowship under Award 2138121. A. Bishop thanks the Belgian American Education Foundation for their Graduate Fellowship support.
A. Vieregg, C. Deaconu, and P. Windischhofer thank the NSF for Award 2411662.

\end{document}